\documentclass[1p]{elsarticle}
\usepackage{graphicx}
\usepackage{dcolumn}
\usepackage{bm}
\usepackage{amsmath}
\usepackage{amssymb}
\begin{document}

\title{Block-determinant formalism for an action of a multi-terminal scatterer}
\author{Yuli V. Nazarov }
\ead{y.v.nazarov@tudelft.nl}
\address{ Kavli Institute of NanoScience, Faculty of Applied Sciences, Delft University of Technology, Lorentzweg 1 2628 CJ Delft,
The Netherlands}

\begin{abstract}
The scattering theory of electron transport allows for a compact and powerful description in terms of $\check{g}^2 = 1$ Green functions, so-called circuit theory of quantum transport. A scatterer in the theory is characterized by an action, most generally a Keldysh one, that can be further used as a building bock of theories describing statistics of electron transport, superconducting correlations, time-dependent and interaction effects. The action is usually used in the form suitable for a two-terminal scatterer.
  
Here we provide a comprehensive derivation of 
a more general form of the action that is especially suitable and convenient for general multi-terminal scatterers. The action is expressed as a determinant of a block of the scattering matrix obtained by projection on the positive eigenvalues of the Green functions characterizing the reservoirs. We start with traditional Green function formalism introducing $\check{g}^2 = 1$ matrices and  give a first  example of multi-terminal counting statistics. Further we consider one-dimensional channels and discuss chiral anomaly arising in this context. Generalizing on many channels and superconducting situation, we arrive at the block-determinant relation. We give the necessary elaborative examples reproducing basic results of counting statistics and super-currents in multi-terminal junctions.  
\end{abstract}
\maketitle
\section{Introduction}
The well-established and refined culture of theoretical description of electron transport in bulk solids and heterostructures was based on field-theoretical methods \cite{AGD} and Keldysh Green functions \cite{Landavshiz}. The pioneering works of Landauer and Buttiker \cite{LB1, LB2} that unambiguously related electron transport and coherent scattering in micro-contacts  have been regarded with suspicion: the genuine simplicity of their approach looked as a barbaric intrusion to a sophisticated domain. It took time to appreciate the idea that the electron resistance is in fact scattering. Once the appreciation of this revolutionary idea was in place, a fast research progress has revealed many facets of the universality of scattering approach and its relevance in the areas where its applicability was not at all obvious. The electic noise was understood in terms of scattering \cite{NoiseButtiker, NoiseLesovik}. The sophistication came back when a state-of-the-art quantum calculation \cite{LesovikLevitov} has demonstrated that the whole statistics of electron transport is defined by the scattering matrix. The approach has been applied to superconducting contacts.

This also initiated research that combined Green function approaches with the notions of scattering and discrete elements giving rise to a bunch of so-called quantum circuit theories \cite{QT} that are indispensable for accessing full counting statistics, transmission distribution in complex scatterers, superconducting and spin transport in nanostructures. A starting point of this research was actually the paper of Buttiker and Beenakker about the noise in diffusive connector \cite{BB} that seemed wrong to the author and thus motivated him to prove the statement on more solid grounds. Quantum circuit theories posses a remarkable degree of universality. A two-terminal scatterer in this approach is always described by an action 
\begin{equation}
{\cal S} = \frac{1}{2} \sum_{n} {\rm Tr} \left[\ln\left(1+T_n \frac{\check{G}_1 \check{G}_2 + \check{G}_2 \check{G}_1 -2}{4} \right)\right] 
\label{eq:twoterminal}
\end{equation}
where $n$ labels transport channels of the scatterer, $T_n$ are corresponding transmission coefficients, while the matrices $\check{G}_{1,2}$ characterize the states of the leads, steam from the Green functions and satisfy $\check{G}_{1,2}^2 =1$. The matrix structure as well as the role of the action conforms a concrete  situation from the great variety where the relation can be applied. In case of circuit theory of transmission distribution \cite{Ohm} $\check{G}$ is a single-parametric $2\times2$ matrix and ${\cal S}$ is function to be minimized while in a theory encompassing interplay of Coulomb interaction and disorder \cite{Efetov} $\check{G}$ may represent supersymmetric $\sigma$-model quantum fields, matrix structure includes time indexes and the action is a part of a path integral weight.

In this article, we address the 
generalization of Eq. \ref{eq:twoterminal} to the case of multiple terminals. Such generalization has hardly been discussed in the literature. One of the reasons for this is the fact that in a (quantum) circuit theory a multi-terminal scatterer can be readily modelled with two-terminal ones and at least a single node connected to the leads by means of these two-terminal scatterers. For instance, this is a way to multi-terminal counting statistics.\cite{NazarovBagrets}  However, such approach is not general. On mean-field level, it disregards random phase factors accumulated in the course of the scattering in the node. The statistics of these random phase factors can in principle be obtained if going beyond the mean-field level. However, this does not give an action for a concrete realization of these phase factors. Such action is especially important in the context of recent discovery of non-trivial topological phenomena in multi-terminal superconducting junctions \cite{MeGrenoble}.

A proposal for such generalization has been made in \cite{Izak} in the context of understanding Fermi Edge singularity. In this paper, we provide a full and comprehensive derivation of this relation starting from the common textbook Green functions and extend it to the case of superconducting terminals. The result is expressed as a determinant of a block of the scattering matrix obtained by projection on the positive eigenvalues of the Green functions characterizing the reservoirs and is given by Eqs. \ref{eq:answer},\ref{eq:answer-sup}. 

The structure of the article is as follows. In Section \ref{sec:green} we discuss the Keldysh Green functions, its extension to counting statistics and give a single-state multiterminal example. In Section \ref{sec:1dchannel} we consider a transport channel connecting two reservoirs without scattering, compute Green functions and the action, recognize and heal a dangerous chiral anomaly. The generalization to many channels and scattering comes in the Section \ref{sec:scattering}. We introduce the superconducting reservoirs in Section  \ref{sec:supres} and generalize the action to this case in Section \ref{sec:super}. Further we elaborate on two basic examples for the block-determinant formula obtained. In Section \ref{sec:normalFCS} we derive the full counting statistics for multi-terminal transport of the normal electrons. In Section \ref{sec:andreev} we perform the projection in superconducting case, consider the ground state energy of the junction and derive a usefull formula for non-stationary superconducting current. We conclude in Section \ref{eq:conclusions}.

\section{Green functions: general}
\label{sec:green}
We start our considerations with conventional definition \cite{Landavshiz} of Keldysh Green functions in terms of averages of fermion creation-annihilation operators $\Psi(t,X)$, $X$ being an element of a Hilbert space (for instance, space coordinate)
\begin{align}
i \check{G}(X_1,X_2) = i \left[\begin{array}{cc} G^{++} & G^{+-} \cr G^{-+} & G^{--}\end{array}\right] = \nonumber \\ = \langle\Psi_1 \Psi^\dagger_2\rangle \left[\begin{array}{cc} \Theta_- & 1 \cr 0 & \Theta_+ \end{array}\right] 
- \langle\Psi^\dagger_2\Psi_1 \rangle \left[\begin{array}{cc} \Theta_+ & 0 \cr 1 & \Theta_- \end{array}\right];\; \Theta_{\pm} \equiv \Theta(\pm(t_1-t_2))
\end{align} 
Here $\Psi_{1,2} \equiv \Psi(X_{1,2})$ and "check" denotes the matrix structure in  the Keldysh index $i=\pm$.
Let us specify to a general stationary  nonequilibrium state where the density matrix is diagonal in the space of energy levels $k$ and filling factor of this level is $f_k$ Since $\Psi(t) = \exp(-i \epsilon_k t)\Psi(0)$, the Green function is diagonal in the levels and reads
\begin{equation}
\label{eq:timeG}
i\check{G}_k (t_1,t_2) = \exp(i\epsilon_k(t_2-t_1)) \left( \left[\begin{array}{cc} \Theta(t_2-t_1) & 1 \cr 0 & \Theta(t_1-t_2) \end{array}\right] -f_k \left[\begin{array}{cc} 1 & 1 \cr 1 & 1 \end{array}\right] \right).
\end{equation}
To get it in the energy representation, $\check{G}(\epsilon) \equiv \int dt e^{i\epsilon t} \check{G}(t,0)$
we note that 
\begin{eqnarray}
\int dt e^{i(\epsilon-\epsilon_k)t -\delta t} \Theta(t) &=& \frac{i}{\epsilon +i\delta -\epsilon_k} \equiv i G_R;\; {\rm Im} G_R = -i \pi \delta(\epsilon-\epsilon_k) \\
\int dt e^{i(\epsilon-\epsilon_k)t -\delta t} \Theta(-t) &=& \frac{-i}{\epsilon -i\delta -\epsilon_k} \equiv -i G_A;\; {\rm Im} G_A = i \pi \delta(\epsilon-\epsilon_k) \\
\int dt e^{i(\epsilon-\epsilon_k)t -\delta t}  &=& 2\pi \delta(\epsilon-\epsilon_k) \equiv i (G_R-G_A) 
\end{eqnarray}
where we have introduced advanced and retarded Green functions not depending on the filling factors. With those, we can represent the Green function as
\begin{equation}
\check{G} = G_R \left[\begin{array}{cc} -f & 1-f \cr -f & 1-f\end{array}\right] + G_A \left[\begin{array}{cc} f-1 & f-1 \cr f & f\end{array}\right]
\end{equation}

For proper description of the reservoirs we need what in old literature is called "Green function in coinciding points", $\check{G}(X,X)$. In fact, this concept has little to do with geometric proximity of the points: rather, it represents a Green function "avegared" over a big number of similar states of a quasicontinuous spectrum of the same energy. Let us generally define it as $\check{G}_{{\rm av}} = \sum_k w_k \check{G}_k$, $w_p$ being some positive weights. The result of such averaging reads  
\begin{equation}
\check{G}_{{\rm av}} = -i \pi \nu  \left[\begin{array}{cc} 1-2f & 2(1-f) \cr -2f & 1-2f\end{array}\right] 
\end{equation}
where $\nu \equiv \sum_k w_k \delta(\epsilon_k -\epsilon)$. For Green function in coinciding points, $\nu$ is the density of states. Its dependence on $\epsilon$ can be disregarded in all important cases. If the filling factors of the levels before the averaging depend on energy only, the filling factor $f(\epsilon)$ in the above relation natually reproduces $f$ of the levels. If not, $f(\epsilon)$ is a weighted average of those and effective filling factor of this group of the states if they are used as a reservoir. 

The common Keldysh technique can be defined through the unitary evolution of the density matrix:
$$
\rho(t) = T\exp( -i H t) \rho(-\infty) \tilde{T}\exp( i H t).
$$
The extended Keldysh technique \cite{QT} is defined through a non-unitary evolution of the pseudo-density matrix with the Hamiltonians $H_{\pm}$ depending on the Keldysh index
\begin{equation}
\rho(t) = T\exp( -i H_- t) \rho(-\infty) \tilde{T}\exp( i H_+ t)
\end{equation}
We define the action ${\cal S}$ in terms of the trace of this pseudo-density matrix after its evolution over a big interval of time ${\cal T}$,
$e^{{\cal S}}={\rm Tr} \rho({\cal T})$. 
Common application of extended Keldysh technique is full counting statistics of electron transfers\cite{Kindermann}.
In this case, $H_{\pm} = H \pm \chi/2 I$, $I$ being the operator of current to a certain reservoir and the Fourier transform of $e^{{\cal S}}$ with respect to $\chi$ gives the probabilities of transferring $N$ electrons during the time interval ${\cal T}$,
$$
P_N = \int_0^{2\pi} \frac{d\chi}{2\pi} e^{i\chi N} e^{{\cal S}(\chi)}.
$$

We will be interested in variations of the action. Let us assume that the Hamiltonians have been changed by a little addition $
H_{\pm} \to H_{pm} + \sum_{ab} h^{\pm}_{ab}(t) \Psi^\dagger_a \Psi_b$. The corresponding variation of the action in the limit of small $h$ is expressed in terms of the Green functions,
\begin{eqnarray}
\delta {\cal S} = -i \int dt h^-_{ab}(t) \langle \Psi^\dagger_a(t) \Psi_b(t) \rangle_{-} +i\int dt h^+_{ab}(t) \langle \Psi^\dagger_a(t) \Psi_b(t) \rangle_{+} = \nonumber \\
= \int dt  \left(- h^-_{ab}(t) G^{ba}_{--}(t+0,t) + h^+_{ab}(t) G^{ba}_{++}(t-0,t) \right)
\end{eqnarray}

The sign difference for $\pm$ in above equation is inconveniently annoying. Also, the evolution equation for $\check G$ has this inconvenient sign difference, 
\begin{equation}
\left(\epsilon - \hat{H}\right) \check{G} = - \tau_z .
\end{equation}
To avoid this nuisance, we better redefine $\check{G}_{new} =  - \check{G}_{old} \tau_z$ departing from the conventional definition.
With this redefinition,
\begin{equation}
\check{G} = G_R \left[\begin{array}{cc} f & 1-f\cr f & 1-f\end{array}\right] + G_A \left[\begin{array}{cc} 1-f & f-1 \cr -f & f\end{array}\right]
\end{equation}
and the averaged function can be presented as
\begin{equation}
\label{eq:averaged}
\check{G}_{{\rm av}} = - i \pi \nu \check{g}; \; \check{g} = \left[\begin{array}{cc} 2f-1 & 2(1-f) \cr 2f & 1-2f \end{array}\right].
\end{equation}
Importantly, $\check{g}^2 = 1$. 
The variation of the action becomes
\begin{equation}
\delta {\cal S} =- \int dt  \left( h^-_{ab}(t) G^{ba}_{--}(t+0,t) + h^+_{ab}(t) G^{ba}_{++}(t-0,t) \right) = - {\rm Tr} \left[\check{h} \check{G}\right] 
\end{equation}
where the trace in the last equality includes everything: the Hilbert space, Keldysh index and time.

With this, we are ready to discuss the reservoirs.  We do it in the context of a single level $k$. We take into account the transition matrix elements $t_{kp}$ from a state $k$ to a reservoir whose states are labelled by $p$. This brings self-energy to the equation for $\check{G}_k$
\begin{equation}
\check{\Sigma} =  \sum_p |t_{kp}|^2 \check{G}_p = -i (\Gamma/2) \check{g}; \Gamma = 2\pi \sum_p |t_{pk}|^2 \delta(\epsilon_p -\epsilon)
\end{equation}
where we have averaged $\check G_p$ over the states, disregarded the $\epsilon$ dependence of $\Gamma$ and the real part of $\Sigma$ since the latter only causes an unimportant shift of $\epsilon_k$.

The Green function of the state $k$ then reads
\begin{equation}
\check{G}_k = \frac{1}{\epsilon -\epsilon_k - \check\Sigma} = 
\frac{1}{\epsilon -\epsilon_k + i (\Gamma/2) \check{g}} = \sum_{\pm} \frac{1\pm\check{g}}{2} \frac{1}{\epsilon -\epsilon_k \pm i (\Gamma/2) } 
\end{equation} 
If we take many states connected to the same reservoir, and average the Green functions over these states, we get ${G}_{{\rm av}} = -i\pi \nu \check{g}$. This implies that the reservoir is reproduced and is completely characterized by the matrix $\check{g}$.

Let us concentrate on the action. The formula for the variation can be rewritten
as 
\begin{equation}
\label{eq:dSdSigma}
\delta {\cal S} = - {\rm Tr} \left[\delta (\check{\Sigma}) \check{G_k}\right],
\end{equation}
from this it follows that 
\begin{equation}
{\cal S} = -{\rm Tr} \left[ \ln \check{G_k}\right]
\end{equation}
this makes sense: $e^{{\cal S}}$ is the determinant of the matrix determining the quadratic Grassman action of the creation and annihilation operators.

Let us connect the level $k$ to a number of reservoirs labelled by $i$ and consider full counting statistics of charge transfers between the reservoirs.
Since there are several reservoirs, the self-energy is contributed by all of them,
\begin{equation}
\Sigma = -i \sum_i (\Gamma_i/2) \check{g}_i
\end{equation}
$\check\Sigma$ is a $2\times2$ traceless matrix, $\check\Sigma^2 = - \Sigma^2$,
the scalar $\Sigma^2$ equals
$$
\Sigma^2 = \frac{1}{4}\left(\sum_i\Gamma^2_i + \sum_{i>j} \Gamma_i \Gamma_j (\check{g}_i \check{g}_j+\check{g}_j \check{g}_j) \right)
$$
\begin{equation}
{\cal S} = \int \frac{d\epsilon}{2\pi} \ln \left( (\epsilon -\epsilon_k)^2 +\Sigma^2\right)
\end{equation}

To incorporate full counting statistics, we transform $\check{g}$ with proper counting fields $\chi_i$, $\check{g} \to e^{i\chi/2 \tau_z} \check{g} e^{-i\chi/2 \tau_z}$, so that the  $\check{g}$ of a given reservoirt becomes $$
\check g = \left[\begin{array}{cc} 2f-1 & 2(1-f)e^{i\chi} \cr 2f e^{-i \chi} & 2f-1\end{array}\right] 
$$
with this, 
$$
\check{g}_i \check{g}_j+\check{g}_j \check{g}_j = 2+4(f_i(1-f_j) (e^{i(\chi_j-\chi_i)}-1)+f_j(1-f_i) (e^{i(\chi_i-\chi_j)}) 
$$

Theoretically, the action should vanish in the limit of $\chi \to 0$. However, this result may be difficult to reproduce without a tedious calculation. It is more  convenient to compute the action allowing some freedom in the insignificant factors that bring extra $\chi$-independent terms and than to substract the value of the computed action at $\chi=0$ to correct for this. This is what we do now
to arrive at 
\begin{equation}
{\cal S} = {\cal T} \int \frac{d\epsilon}{2\pi} \ln\left(1+\sum_{i\ne j} T_{ij} f_i(1-f_j) (e^{i(\chi_j-\chi_i)}-1) \right). 
\end{equation}
This describes statistics of scattering of the electrons from one  reservoir to another trought the level $k$ with energy-dependent transmission coefficients $T_{ij}$,
$$
T_{ij} = \frac{\Gamma_i \Gamma_j}{(\epsilon -\epsilon_k)^2 + (\sum_j \Gamma_j/2)^2}.$$

 To remind, $\cal{T}$ is a time interval at which the statistics is acquired.
If $\Gamma$ is much smaller than the scale at which $f$ is changing (this is usually a temperature),
this expression can be integrated over the energy. 
\begin{equation}
{\cal S} = \sqrt{(\sum_j \Gamma_j/2)^2 +\sum_{i\ne j} \Gamma_i \Gamma_j f_i(1-f_j) (e^{i(\chi_j-\chi_i)}-1) } - \sum_j \Gamma_j/2.
\end{equation}
 The $f_i(\epsilon)$ in the above expression are taken at $\epsilon = \epsilon_k$.
 
In fact, we have obtained a simplest model of a multi-terminal scattering: that generated by transitions through a single resonant level. It is not the most general form of the scattering matrix. To consider this one, we need to turn to Green functions of transport channels. Those can be conventionally modelled assuming a 1d spacial dependence \cite{LB1,LB2}. 

\section{Single one-dimensional channel}
\label{sec:1dchannel}
Let us consider a 1d channel with the Hamiltonian
\begin{equation}
\hat{H} = -i v \partial_x 
\end{equation}
sign of the velocity $v$ is important here: at the moment, we assume it is positive. Similarly to what we have done for a single level, let us consider weak coupling of the channel states to those of a reservoir. This results in a self-energy part $\check \Sigma = (-i/2\tau) \check{g}$, where $\tau$ is a typical time of escape from the channel to the reservoir, $\check{g}$ characterizing the reservoir.
Green function of the electrons in the channel satisfies the differential equation 
\begin{equation}
\left(\epsilon + i v \partial_x  + (i/2\tau) \check{g} \right)\check{G}(x,x') = \delta(x-x')
\end{equation}
The differential equation with respect to another coordinate reads:
\begin{equation}
- i v \partial_x\check{G}(x,x') + \check{G}(x,x')\left(\epsilon   + (i/2\tau) \check{g} \right) = \delta(x-x')
\end{equation}
We can solve it in $x$ deriving
\begin{equation}
\check{G}(x_1,x_2) = \check{R}(x_1 -x'_1) \check{G}(x_1,x_2) \check{R}^{-1}(x_2 -x'_2); \check{R} \equiv \exp\left( - x \frac{\check{g}}{2\tau v} \right)
\end{equation}
Since $\check{g}^2 =1$, for any $A$
$$
\exp\left( A {\check{g}} \right) = \frac{1+\check{g}}{2} \exp(A) + \frac{1-\check{g}}{2} \exp(-A).
$$

Let us complicate the situation and consider a channel connected to two reservoirs. To this extent, let us introduce a simple $x$-dependence of the reservoir Green function: at $x<0$, $\check{g}= \check{g}_1$, and at $x>0$ $\check{g}= \check{g}_2$. Let us concentrate on the Green function in coinciding points, $\check{G}(x) =\check{G}(x,x')$. For a convenient normalization, it is prudent to rewrite it as $\check{G} = -i\pi\nu \check{Q} = -i(1/2v) \check{Q}$. Actually, $\check{Q}(x,x')$ satisfies the equation
\begin{equation}
\left(-i \epsilon/v + \partial_x + \frac{1}{2\tau v} \check{g} \right) \check{Q}(x,x') = \delta(x-x') 
\end{equation}
We can express $\check{Q}(x)$ in terms of $\check{Q}(0)$ at the "boundary",
\begin{eqnarray}
x<0: & \; \check{Q}(x) &= \sum_{\pm} e^{\pm\frac{x}{\tau v}} \frac{1\mp\check{g}_1}{2} \check{Q}(0) \frac{1\pm\check{g_1}}{2}  - \frac{1}{2} \left(\check{Q}(0)-\check{g}_1 \check{Q}(0)\check{g}_1 \right) \\
x<0: & \; \check{Q}(x) &= \sum_{\pm} e^{\pm\frac{x}{\tau v}} \frac{1\mp\check{g}_2}{2} \check{Q}(0) \frac{1\pm\check{g_2}}{2} + \frac{1}{2} \left(\check{Q}(0) - \check{g}_2 \check{Q}(0)\check{g}_2 \right) 
\end{eqnarray}
To determine $\check{Q}(0)$, we note that for the solution not to diverge at the infinities we need to require
\begin{equation}
(1+\check{g}_1)\check{Q}(0) (1-\check{g}_1) = 0;\; (1-\check{g}_2) \check{Q}(0) (1+\check{g_2}).
\end{equation}
Besides, by reproducibility of the reservoir, we have to require $\check{Q(x)} \to \check{g}_{1,2}$ at $x \to \mp \infty$. With this,
\begin{equation} \label{eq:Q0}
\check{Q(0)} = (1-\check{g}_1)\frac{1}{\check{g}_1 +\check{g}_2} + \frac{1}{\check{g}_1 +\check{g}_2}(1+\check{g}_1) = (1+\check{g}_2)\frac{1}{\check{g}_1 +\check{g}_2} + \frac{1}{\check{g}_1 +\check{g}_2}(1-\check{g}_2).
\end{equation} 
Actually, $\check{Q}(x,x')$ has a discontinuity at $x\to x'$. One shows that $\check{Q}(x\pm 0,x) = \pm \check{1} +\check{Q}(x)$.

Let us compute the action. We vary the Green function of the first reservoir,
\begin{eqnarray*}
\delta{\cal S} = - \int_{-\infty}^0 dx {\rm Tr}\left[\delta(\Sigma(x)) \check{G}(x,x)\right] \\
= \frac{1}{4\tau v} \int_{-\infty}^0 dx {\rm Tr}\left[\delta(\check{g_1})(x) \check{Q}(x)\right]\\
\end{eqnarray*} 
since ${\rm Tr} \left[\delta(\check{g}) \check{g} \right] =0$ for $\check{g}^2=1$ matrices, the integral is accumulated at distances of the order of 
$1/v\tau$ from the interface where $\check{Q}(0)$ changes into $\check{G}$,
\begin{equation}
\delta{\cal S} = \frac{1}{4\tau v} \int_{-\infty}^0 dx {\rm Tr}\left[\delta(\check{g_1}) \check{Q}(0) \right] e^{\frac{x}{\tau v}} = -\frac{1}{4} {\rm Tr}\left[\delta(\check{g_1}) \check{Q}(0) \right]\end{equation} 
Substituting  (\ref{eq:Q0}) yields 
\begin{equation}
\label{eq:deltaaction}
\delta{\cal S} = \frac{1}{2} {\rm Tr}\left[\delta(\check{g}_1)(1-\check{g}_1)\frac{1}{\check{g}_1 +\check{g}_2}  \right]
\end{equation} 
Now we need to integrate over $\check{g}_1$ to determine the action.
Eventually, this does not quite work. 
To make this explicit, let us do this for $2 \times 2$ matrices $\check{g}_{1,2}$. This is general case, since $\check{g}_{1,2}$ can always be presented in the form of $2\times2$ blocks in the space of doubly degenerate eigenvalues of $\check{g}_{1}\check{g}_{2} +\check{g}_{2}\check{g}_{1}$. We substitute $\check{g} = \vec{g} \cdot \vec{\check{\sigma}}$ to obtain 
\begin{eqnarray*}
\delta{\cal S} = \delta {\cal S}_n +\delta {\cal S}_a; \\
\delta {\cal S}_n =  \frac{(\delta \vec{g}_1, \vec{g}_2)}{(\vec{g}_1+\vec{g}_2)^2} = \frac{1}{2} \delta(\ln(1+(\vec{g}_1,\vec{g}_2));\\
\delta {\cal S}_a = -\frac{1}{2} \frac{(\delta \vec{g}_1, \vec{g}_1,\vec{g}_2)}{1+(\vec{g}_1,\vec{g}_2)}
\end{eqnarray*} 
The variation of ${\cal S}_n$ can be easily integrated. For general $\check{g}$,
\begin{equation}
{\cal S}_n = \frac{1}{2} {\rm Tr}\left[\ln \frac{\check{g}_1 +\check{g}_2}{2}\right]
\end{equation}  
where we have added factor of 2 to make sure ${\cal S}_n =0$ if both reservoirs are identical.
 
As a matter of fact, the term $\delta{\cal S}_a$ in the action cannot be integrated.  Indeed, if we set $\check{g}_2 \parallel z$, $\check{g}_1 = (\cos \theta, \sin \theta \sin \phi, \sin \theta \cos \phi)$, we obtain
$$
\delta {S}_a = -\delta\phi \ \frac{\sin^2\theta}{2(1+\cos\theta)},
$$
that is, $\partial_\phi \partial_\theta {\cal S} \ne \partial_\phi \partial_\theta {\cal S} $.

This signals the breakdown of the perturbation theory in $\check{\Sigma}$ and is in fact a manifestation of chiral anomaly. A chiral channel between two different reservoirs is not well-defined, for instance, it would provide an infinite current between them. One can think of a less invasive configuration, for instance, $\check{g} =\check{g}_1$ at $x \to \pm \infty$ and $\check{g}=\check{g}_2$ in the  finite but long interval of  $x$ In this case, the two boundaries between regions with $\check{g}_{1,2}$  produce the $\cal{S}_a$ of opposite sign, so that the anomalous part cancels and the action is integrable. This is because the channel starts and ends up in identical reservoirs. 
An equivalent picture includes two channels of opposite velocities between two reservoirs. The ${\cal S}_a$ terms cancel and ${\cal S}_n$ add resulting in
\begin{equation}
{\cal S} = {\rm Tr} \left[ \ln \frac{\check{g}_1 +\check{g}_2}{2}\right]
\end{equation}
If we substitute the reservoirs with counting field, we reproduce the known expression for counting statistics of the quantum point contact of ideal transparency,
\begin{equation}
{\cal S} = \int \frac{d\epsilon}{2\pi} \ln \left( 1+ f_1(1-f_2)e^{i\chi} + f_2(1-f_1)e^{-i\chi}\right)
\end{equation}

This eventually sets the model in use: further we assume that the channels are grouped in pairs of opposite velocity, and the members of the pair share the same reservoir.

\section{Scattering}
\label{sec:scattering}
Let us have many (pairs) of channels and describe scattering between those. Without loosing the generality we assume  that all channels
are at $x<0$ and the wave function amplitudes at $x=0$ are related by scattering matrix,
$$
\psi^{out}_i = \sum_j S_{ij}  \psi^{in}_j
$$
We incorporate the index of the channel into the "check" index. To simplify the formulas, we assume 

Let us compute the Green function of the incoming electrons at a point $x'$. We define $\check{Q}^{\pm} \equiv \check{Q}(x'\pm 0, x')$. For the solution not to diverge at $x \to -\infty$, it should satisfy  
$$
\left(1+\check{g}\right) \check{Q}^{-}=0. 
$$  
At $x >x'$, the $Q$ evolves to $\check{Q}^{+}(0,x') = \check{R}(-x')\check{Q}^+$.
Further, it scatters to the outgoing channels where the amplitudes should satisfy 
the condition with changed sign of $\check{g}_1$. This gives
$$
\left(1-\check{g}\right)\check{S}\check{R}(-x') \check{Q}^{+} =0 
$$  
We multiply this equation by $\check{S}^{-1}$ from the left to arrive at 
\begin{equation}
\left(1-\check{g}_3\right) \check{R}(-x') \check{Q}^{+} =0; \; \check{g}_3 \equiv \check{S}^{-1} \check{g} \check{S}.
\end{equation}   
We substitute $\check{Q}^{\pm} = R(x')\check{q}^{\pm} R^{-1}(x')$ to arrive at 
\begin{equation}
\label{eq:q} 
\left(1+\check{g}\right) \check{q}^-=0; \left(1-\check{g}_3\right) \check{q}^-=0; q^{+}-\check{g}^{-} = 2.
\end{equation}
Substituting $\check{q}^{\pm} = \check{q} \pm 1$, we arrive at the answer similar to Eq. \ref{eq:Q0},
\begin{equation}
\check{q} = (1-\check{g})\frac{1}{\check{g} +\check{g}_3} + \frac{1}{\check{g} +\check{g}_3}(1+\check{g}) = (1+\check{g}_3)\frac{1}{\check{g} +\check{g}_3} + \frac{1}{\check{g} +\check{g}_3}(1-\check{g}_3).
\end{equation} 
Repeating the steps that lead to Eq. \ref{eq:deltaaction}, we arrive at 
\begin{equation}
\delta{\cal S} = \frac{1}{2} {\rm Tr}\left[\delta(\check{g})(1-\check{g})\frac{1}{\check{g} +\check{g}_3}  \right]
\end{equation}
Now we need to compute the contribution of the outgoing channels. There are sign differences in the equations indicating the opposite sign of the velocity. For instance, Eq.\ref{eq:q} becomes
\begin{equation}
\label{eq:q} 
\left(1+\check{g}\right) \check{q}^-=0; \left(1-\check{g}_4\right) \check{q}^-=0; q^{+}-\check{g}^{-} = 2; \; \check{g}_4 \equiv \check{S}\check{g}\check{S}^{-1}
\end{equation}
As a result of this, the contribution of the outgoing channels reads 
\begin{equation}
\delta{\cal S} = \frac{1}{2} {\rm Tr}\left[\delta(\check{g})(1+\check{g})\frac{1}{\check{g} +\check{g}_4}  \right]
\end{equation}
We bring both contributions together in the following form
\begin{equation}
\label{eq:variation} 
\delta{\cal S}= \frac{1}{2} {\rm Tr} \left[\left(\check{S} \delta(\check{g})(1+\check{g}) + \delta(\check{g}) (1-\check{g}) \check{S} \right)\frac{1}{\check{g} \check{S} + \check{S}\check{g}}\right]
\end{equation}
 
Now it is time to integrate. The integration is less trivial than one could expect. The practical way is to do it in the basis where $\check{g}$ is diagonal. The positive and negative eigenvalues of the matrix define a block structure:
\begin{equation}
\check{g} = \left[\begin{array}{cc} 1 & 0 \cr 0 &-1\end{array}\right]; \; \check{S} = \left[\begin{array}{cc} \check{S}_{11} & \check{S}_{12} \cr \check{S}_{21} & \check{S}_{22}\end{array}\right]
\end{equation}
The action after the integration (see Appendix for details) is expressed in terms of the determinant of the upper left block of $\check{S}$,
\begin{equation}
\label{eq:answer}
{\cal S} = {\rm Tr}\left[ \ln \check{S}_{11}\right] = \ln {\rm det}\left(\check{S}_{11}\right).
\end{equation}
This formula can be presented in a basis-invariant form in a variety of ways, the most logical one is the following:
\begin{equation}
{\cal S} = {\rm Tr}\left[ \ln \left( \frac{1-\check{g}}{2} + \frac{1+\check{g}}{2}\check{S}\frac{1+\check{g}}{2}\right)\right] 
\end{equation}
where the first term is the projector on the down right block, and the second one is the projection of $\check{S}$ on the upper left block.

\section{Superconducting reservoirs}
\label{sec:supres}
Let us generalize the Keldysh Green function of normal electrons
\begin{eqnarray}
i \check{G}(X_1,X_2) = i \left[\begin{array}{cc} - G^{++} & G^{+-} \cr - G^{-+} & G^{--}\end{array}\right] = \langle\Psi_1 \Psi^\dagger_2\rangle \left[\begin{array}{cc}-\Theta(t_2-t_1) & 1 \cr 0 & \Theta(t_1-t_2) \end{array}\right]  \nonumber \\
- \langle\Psi^\dagger_2\Psi_1 \rangle \left[\begin{array}{cc} -\Theta(t_1-t_2) & 0 \cr -1 & \Theta(t_2-t_1) \end{array}\right] 
\end{eqnarray} 
to superconducting state. We label electron operators with  $\alpha=\pm 1, \Psi_1 = \Psi, \Psi_{-1} = \Psi^{\dagger}$ to introduce the Green function with Nambu indices
\begin{eqnarray}
i \check{G}_{\alpha\beta} (X_1,X_2) = \langle\Psi_{1,\alpha} \Psi_{2,-\beta} \rangle \left[\begin{array}{cc} -\Theta(t_2-t_1) & 1 \cr 0 & \Theta(t_1-t_2) \end{array}\right] \nonumber \\
- \langle\Psi_{2,\beta}\Psi_{1,\alpha} \rangle \left[\begin{array}{cc} -\Theta(t_1-t_2) & 0 \cr -1 & \Theta(t_2-t_1) \end{array}\right] 
\end{eqnarray}  
Given a superconducting Hamiltonian (where $\hat{H}$ is Hermitian and $\hat{\Delta} = - \hat{\Delta}^T$)
\begin{equation}
\hat{{\cal H}} = \sum_{k,l} \Psi^{\dagger}_k H_{kl} \Psi_l + \frac{1}{2} \Psi^{\dagger}_k \Delta_{kl} \Psi^{\dagger}_l   + \frac{1}{2} \Psi_l \Delta^*_{kl} \Psi^{\dagger}_k 
\end{equation}
the time-dependence of $\Psi$ is given by the Bogolyubov-deGennes "Hamiltonian" ${\cal H}_{{\rm BdG}}$,
\begin{equation}
i\frac{\partial}{\partial t} \left[\begin{array}{c} \Psi_1 \cr \Psi_{-1} \end{array} \right] = {\cal H}_{{\rm BdG}}\left[\begin{array}{c} \Psi_1 \cr \Psi_{-1} \end{array} \right]; \; {\cal H}_{{\rm BdG}} \equiv \left[\begin{array}{cc} \hat{H} & \hat{\Delta} \cr \hat{\Delta}^{\dagger} & - \hat{H}^{T}  \end{array} \right]. 
\end{equation}
With this, in energy representation the Green function satisfies 
\begin{equation}
\left(\epsilon - {\cal H}_{{\rm BdG}}\right)\check{G} = \check{1}.
\end{equation}
This is not convenient since the usual potential comes to this equation in the form $U(x)\eta_z$, $\vec{\eta}$ being the Pauli matrices in Nambu space. To circumvent this, we redefine $\check{G}_{new} = \check{G}\eta_z$. The redefined function satisfies 
\begin{equation}
\left(\epsilon \eta_z - \bar{{\cal H}}_{{\rm BdG}}\right)\check{G} = \check{1}; \; \bar{{\cal H}}_{{\rm BdG}} = \equiv \left[\begin{array}{cc} \hat{H} & \hat{\Delta} \cr -\hat{\Delta}^{\dagger} & - \hat{H}^{T}  \end{array} \right]
\end{equation}
We can equilidate the standard spin structure of $\hat{\Delta} = i\sigma_{y} \hat{\Delta}$ by making unitary transformation of $\check{G}$, $\bar{{\cal H}}_{{\rm BdG}}$ with the matrix $((1+\eta_z)+\sigma_y(1-\eta_{z}))/2$ to arrive at
\begin{equation}
\bar{{\cal H}}_{{\rm BdG}} = \equiv \left[\begin{array}{cc} \hat{H} & -i \hat{\Delta} \cr i\hat{\Delta}^{\dagger} & \hat{\bar{H}} \end{array} \right],
\end{equation} 
where $\hat{\bar{H}}$ is the time-reversed $\hat{H}$.  
Let us compute the Green function of a normal metal averaged over the states. It has a block structure  in Nambu space: the upper block is the same as evaluated in (\ref{eq:averaged}), while the lower block is obtained by replacing $\Psi^{\dagger} \leftrightarrow \Psi$.  For a single state $k$, it therefore reads (c.f. Eq.\ref{eq:timeG}) 
\begin{eqnarray}
\label{eq:timeG}
i\check{G}^{down}_k (t_1,t_2) = \\
- \exp(-i\epsilon_k(t_2-t_1)) \left( \left[\begin{array}{cc} -\Theta(t_2-t_1) & 1 \cr 0 & \Theta(t_1-t_2) \end{array}\right] -(1-f_k) \left[\begin{array}{cc} -1 & -1 \cr 1 & 1 \end{array}\right] \right).
\end{eqnarray} 
In energy representation, it becomes
\begin{equation}
- \check{G}^{down} = G_R \left[\begin{array}{cc} 1-f & f \cr 1-f & f\end{array}\right] + G_A \left[\begin{array}{cc} f & -f \cr -(1-f) & (1-f)\end{array}\right]; \; G^{A,R} \equiv \frac{1}{\epsilon \mp i\delta +\epsilon_k}.
\end{equation}
We notice that upon averaging the filling factor is taken at $-\epsilon$. With  this, both blocks are presented as
\begin{eqnarray}
\label{eq:normal_reservoir}
\check{G} = - i \pi \nu  \left( \frac{1+\eta_z}{2}  \check{F}_+ + \frac{\eta_z-1}{2}\right)\check{F}_- \equiv -i\pi \nu \check{g}_N; \\
 \check{F}_+ = \left[\begin{array}{cc} 2f(\epsilon)-1 & 2(1-f(\epsilon)) \cr 2f(\epsilon) & 1-2f(\epsilon) \end{array}\right];\; \check{F}_- = \left[\begin{array}{cc} 1-2f(-\epsilon) & 2 f(-\epsilon) \cr 2(1-f(-\epsilon)) & 2f(-\epsilon)-1 \end{array}\right]
\end{eqnarray}
If $f(\epsilon)=1-f(-\epsilon)$ as it is the case of Fermi distribution at $\mu=0$,$\check{F}_+=\check{F}_-=\check{F}$, $\check{g}_N = \eta_z \check{F}$.

Let us now compute the Green function of a superconductor connected to a normal metal reservoir.
The reservoir produces the self-energy $\check{\Sigma}_N = -i/\tau \check{g}_N$, we will assume $1/\tau \ll \epsilon, \Delta$. The Green function sought is defined by 
\begin{equation}
\left(\epsilon\eta_z + i \Delta \eta_x -\xi - \check{\Sigma}_N \right) \check{G} = \check{1}.
\end{equation}
assuming real $\Delta$.
If $f(\epsilon)=1-f(-\epsilon)$, the matrix can be diagonalized  in Keldysh structure separately from the Nambu one, and the Green function can be presented with the projectors on the blocks of $\check{F}$,
\begin{equation}
\label{eq:GF}
\check{G} = \frac{1+\check{F}}{2} \check{G}_R + \frac{1-\check{F}}{2} \check{G}_A; \; \check{G}_{R,A} = \frac{1}{(\epsilon \pm i\delta)\eta_z +\Delta \eta_x - \xi}.
\end{equation}
In general case, the solution is more involved. Let us note that if $|\epsilon|<|\Delta|$, $\check{G}^R = \check{G}^A$ and the Green function does not depend on $f$: there is no density of states for such energies and nothing is to be eqilibrated with the normal reservoir. In opposite case, the matrix $\check{E} \equiv \epsilon \eta_z +\Delta \eta_x$ can be presented as $\check{E} = E \check{\epsilon}, \check{\epsilon}^2 =1$. It is convenient to choose the signs in such a way that ${\rm Tr}(\eta_z \check{\epsilon}) \equiv \nu_{B}$ is always positive. In this case, $\nu_{B} = |\epsilon|/\sqrt{\epsilon^2 -\Delta^2}$ is the normalization factor of BCS density of states.  The advanced and retarded Green functions read
\begin{equation}
\check{G}^{R,A} = \frac{1}{\check{E} \pm i \eta_z \delta - \xi} = \frac{1}{E \pm i \delta - \xi} \frac{1+\check{\epsilon}}{2} + \frac{1}{- E \mp i \delta - \xi} \frac{1-\check{\epsilon}}{2} 
\end{equation}
For averaged functions, 
$$
\check{G}^{R,A} = -i \nu \pi \pm \check{\epsilon}.
$$
To compute the Green function, we project the Nambu structure in Eq. \ref{eq:normal_reservoir} onto $\pm$ blocks of $\epsilon$:
\begin{equation}
\check{g}_N \to {\rm Tr}_{{\rm N}}\left[ \frac{1\pm\check{\epsilon}}{2} \check{g}_N \right] = \check{F}_+ \frac{1\pm \nu_B}{2} - \check{F}_- \frac{1\mp\nu_B}{2} \equiv \pm \check{V}_{\pm}
\end{equation}
With this, 
\begin{equation}
\check{G} = \sum_{\sigma,\sigma' =\pm 1} \frac{1}{\sigma E -\xi + i\sigma'\delta} \frac{1+\sigma\check{\epsilon}}{2}\frac{1+\sigma'\sigma\check{v}_\sigma}{2}
\end{equation}
where matrices $v_{\pm}$ are obtained from $\check{V}_{\pm}$ by normalization such that $v_{\pm}^2=1$
For the standard structure not spoiled by any counting fields, $v_{\pm}=\pm\check{V}_{\pm}/\nu_B$, and the matrices $v_{\pm}$ have a structure \ref{eq:Fmatrix} corresponding to the effective filling factors
$$
f^{\pm} = f(\epsilon) \frac{1\pm 1/\nu_B}{2} + (1-f(-\epsilon)) \frac{1\mp 1/\nu_B}{2};\; f^{\pm}(\epsilon) = (1-f^{\mp}(-\epsilon).
$$
The effective filling factors account for charge imbalance in the superconductor that is induced provided 
$f(\epsilon) \ne (1-f(-\epsilon))$.

The average $G$ then reads:
\begin{equation}
i\pi \nu \check{G}= \frac{1+\check{\epsilon}}{2} \check{v}_+ + \frac{\check{\epsilon}-1}{2} \check{v}_-
\end{equation} 

The formulas relating the Green functions and the variation of the action remain the same apart from $1/2$ factor that comes with the trace over Nambu structure and compensates the artificial doubling of fermionic states in Nambu formalism. For instance, Eq.  \ref{eq:dSdSigma} becomes
\begin{equation}
\delta {\cal S} = - \frac{1}{2}{\rm Tr} \left[\delta (\check{\Sigma}) \check{G_k}\right].
\end{equation}
For spin-symmetric case, this factor cancels with the factor of $2$ accounting for spin degeneracy.

\section{Formulation for multi-terminal superconductor}
\label{sec:super}
The Green functions describing the superconducting reservoir can be dealt with in the same way as we deal with the Keldysh functions: we just account for an extra Nambu index. The same holds for possible spin structure or for any other more exotic structure accounting for approximate degeneracy of electron states. 
The only essential modification of the approach developed in Section \ref{sec:scattering} is the Nambu index dependence of the scattering matrix. 
Indeed, the scattering matrix for normal electrons relates annihilation operators
$$
\Psi^{out}_i = \sum_j S_{ij}  \Psi^{in}_j
$$
For creation operators, this is transformed to
$$
\Psi^{\dagger,in}_i = \sum_j S^*_{ij}  \Psi^{\dagger,out}_j.
$$
This defines the Nambu structure of the scattering matrix in conventional terms of "electron" and "hole" scattering,
\begin{equation}
\check{S}_{ij} = \frac{1+\eta_z}{2} s^{e}_{ij} + \frac{1-\eta_z}{2} s^{h}_{ij};\; s^e_{ij} \equiv S_{ij}
\end{equation}
In the simplest case of energy and spin-independent scattering matrix, $s^{h} = (s^{e})^T$. Generally, $s^{h}(\epsilon) = \sigma_y (s^{e}(-\epsilon))^T \sigma_y$.
With this, we reproduce the same formulas as for the normal case:
\begin{equation}
\label{eq:answer-sup}
{\cal S} = \frac{1}{2}{\rm Tr}\left[ \ln \check{S}_{11}\right] = \frac{1}{2}\ln {\rm det}\left(\check{S}_{11}\right).
\end{equation}
or, in basis-invariant form,
\begin{equation}
{\cal S} = \frac{1}{2}{\rm Tr}\left[ \ln \left( \frac{1-\check{g}}{2} + \frac{1+\check{g}}{2}\check{S}\frac{1+\check{g}}{2}\right)\right] 
\end{equation}
As mentioned, the $1/2$ factor compensates for the doubling of states in Nambu representation. It cancels the spin doubling factor for spin-independent scattering.
The formulas \ref{eq:answer}, \ref{eq:answer-sup} provide an ultimately general and compact answer for statistics of electron transport. It is important to understand that the formulas suit perfectly the time-dependent scattering matrices, reservoirs and currents since time is just another index in the "check" matrix structure, and also for time-dependent counting fields.
In this article, we will not go into details of time-dependent situation. Rather, we will elaborate two examples for stationary case when the matrix structure can be resolved separately at each energy.
  
\section{Elaborative example for normal multi-terminal statistics}
\label{sec:normalFCS}
For normal case, we demonstrate the known relations for counting statistics in the limit of long measuring times. 

The Green function in a reservoir connected to the channel $k$ is given by Eq.\ref{eq:averaged}. As any matrix we can present it as an expansion in left and right eigen-vectors, $g_{\alpha \beta} = \sum_{i} g_i \psi^{L}_\alpha \psi^{R}$. The $\check{g}$ has eigenvalues $\pm 1$ and without counting field the eigenvectors read
\begin{equation}
\psi^{L,+} = \left[\begin{array}{c} 1 \cr 1\end {array}\right];\;
\psi^{R,+} = \left[\begin{array}{c} f_k \cr 1-f_k\end {array}\right]; \;
\psi^{L,-} = \left[\begin{array}{c} -(1-f_k) \cr f_k\end {array}\right];\;
\psi^{R,-} = \left[\begin{array}{c} -1 \cr 1\end {array}\right].
\end{equation}
With the counting field $\chi_k$ associated with this particular reservoir, the eigenvectors are changed by unitary transformation 
$\psi^{L,R} \to \exp(i \pm \tau_z \chi_k)\psi^{L,R}$.
To evaluate the action, we need to project onto $+1$ eigenvalue.
Upon the projection, the element of scattering matrix is modified to
\begin{equation}
\label{eq:normalFCS}
\bar{S}_{jk} = S_{jk} \psi^{R,+}_{j,\alpha} \psi^{L,+}_{k,\alpha} = S_{jk}  \left(f_k e^{i(\chi_j-\chi_k)/2} +(1-f_k) e^{i(\chi_k-\chi_j)/2}\right).
\end{equation}
With this, the action for full counting statistics becomes
\begin{equation}
{\cal S} = {\cal T} \int \frac{d\epsilon}{2\pi} \left({\rm ln\ det}\left(\bar{\check{S}}\right) -  {\rm ln\ det}\left(\check{S}\right)\right),
\end{equation} 
where we have added the last term to make sure that ${\cal S}=0$ at $\chi \to 0$. There are many equivalent representations of this formula with matrices that have the same determinant.
To give a useful one, let us transform the matrix $\bar{S}$ with unitary transform $\exp(i\check{\chi}/2)$ and multiply the result with $\check{S}^\dagger$ from the left coming to
\begin{equation}
\check{Q} = \exp(-i\check{\chi}/2) \bar{\check{S}} \exp(i\check{\chi}/2) \check{S}^{\dagger} = 1- \check{f} + \check{f} \exp(-i\check{\chi}) \check{S} \exp(i\check{\chi}) \check{S}^{\dagger}
\end{equation}  
We observe that ${\rm det}(\check{Q}) = {\rm det}(\bar{\check{S}})/{\rm det}(\check{S})$ so that
\begin{equation}
\label{eq:detQ}
{\cal S} = {\cal T} \int \frac{d\epsilon}{2\pi} {\rm ln\ det}\left(\check{Q}\right)
\end{equation}
In this form, this is the same formula as given in seminal work \cite{LesovikLevitov}: actually, Eq. 8 of this work,  that is much less popular than its two-terminal one-channel elaboration. 

For completeness, let us derive this elaboration as well. Let us consider a $2\times2$ scattering matrix  in the basis of the channels $1,2$,
$$
\check{S} =\left[\begin{array}{cc} r_1 & t_{12}\cr t_{21} & r_2 \end{array} \right]
$$
its unitarity implying $|r_1|^2 +|t_{12}|^2 = |r_2|^2 + |t_{21}|^2 =1$, $r_1^*t_{12} +r_2 t_{12}^* =0$.
The matrix $\check{Q}$ thus reads
$$
\check{Q} = \check{1} + \left[\begin{array}{cc} f_1 |t|^2 (\exp(i(\chi_2-\chi_1))-1) & f_1 r_2^* t_{12}(\exp(i(\chi_2-\chi_1))-1)\cr f_2 r^*_1 t_{21}(\exp(-i(\chi_2-\chi_1))-1) & f_2 |t|^2 (\exp(-i(\chi_2-\chi_1))-1) \end{array}\right]
$$
Computing the determinant and substituting the result into \ref{eq:detQ} we arrive at the standart expression for the two-terminal full counting statistics that leads to binomial statistics of the charge transferred,
\begin{equation}
{\cal S} = {\cal T} \int \frac{d\epsilon}{2\pi} {\rm ln}\left(1+ f_1(1-f_2)(\exp(i(\chi_2-\chi_1))-1) + f_2(1-f_1)(\exp(i(\chi_1-\chi_2))-1)\right).
\end{equation} 
  
\section{Elaborative example for a superconducting setup}
\label{sec:andreev}
Let us elaborate on a similar example for a superconducting setup. For simplicity we assume the absence of charge imbalance in the superconducting reservoirs so that each reservoir is described by (c.f. Eq. \ref{eq:GF})
\begin{equation}
\check{G} = \frac{1+\check{F}}{2} \check{g}_R + \frac{1-\check{F}}{2} \check{g}_A;
\end{equation}
To project on the upperleft block of this matrix, we need to find its left and right eigenvectors. Those separate in Nambu (latin index) and Keldysh structure, two independent vectors can be chosen as follows:
\begin{eqnarray}
\Xi^{L}_1 &=& \psi^{L,+}_\alpha R^{L}_a; \check{g}_R R^L = R^L \\
\Xi^{L}_2 &=& \psi^{L,-}_\alpha R^{L}_a; \check{g}_A A^L =  A^L, 
\end{eqnarray} 
and similar for right eigenvectors.
To proceed, it is convenient to parametrize $g_{A,R}$ in the form that corresponds to zero superconducting phase yet suffuciently general one to account for possible complex energy dependence of these functions. The parametrization in terms of complex "phase" $\theta(\epsilon)$ is as follows:
\begin{equation}
\check{g}_R = \frac{1}{\cos\theta} \left[\begin{array}{cc} -i \sin \theta & 1 \cr 1 & i \sin \theta \end{array} \right]; \; \check{g}_A = \frac{1}{\cos\theta^*} \left[\begin{array}{cc} -i \sin \theta^* & 1 \cr 1 & i \sin \theta^* \end{array} \right].
\end{equation}
For pure BCS density of states, $\theta$ is the solution of $\sin\theta = \epsilon/|\Delta|$ in upper half of the complex plane. At $\epsilon <|\Delta|$ it is real and differs by $\pi/2$ shift from the phase of Andreev scattering. \cite{QT} At $|\epsilon| > |\Delta|$, $\theta = \frac{\pi}{2}{\rm sng} \epsilon + i \mu$, where $\mu >0$, ${\rm cosh}\mu = |\epsilon|/\Delta$. With this, the eigenvectors needed become
\begin{equation}
A^L=A^R = \left[\begin{array}{c} e^{-i\theta^*/2} \cr e^{i\theta^*/2}\end {array}\right]\frac{1}{\sqrt{2\cos\theta^*}};\; R^L=R^R = \left[\begin{array}{c} e^{-i\theta/2} \cr e^{i\theta/2}\end {array}\right]\frac{1}{\sqrt{2\cos\theta}}.
\end{equation}

We assign both superconducting phases $\phi_k$ and counting fields $\chi_k$ to the scattering matrix. With this, it becomes
\begin{equation}
\check{S}_c = \frac{1+\eta_z}{2} \left(\frac{1+\tau_z}{2} \check{s}_+ +\frac{1-\tau_z}{2} \check{s}_+\right)+\frac{1-\eta_z}{2} \left( \frac{1+\tau_z}{2} \check{s}^T_+ +\frac{1-\tau_z}{2} \check{s}^T_-\right)
\end{equation}
where $\check{s}_{\pm} = \exp(i\check{\phi}/2 \pm \check{\chi}/2) \check{s} \exp(-i\check{\phi}/2 \mp \check{\chi}/2)$.
We project the matrix $\check{S}_c$ onto the space spanned by the vectors $\Xi_{1,2}$ going from $4\times4$ Nambu-Keldysh structure to a simpler $2\times2$ structure. The result reads ($\check{Z} \equiv \exp(i\check{\theta})$)
\begin{eqnarray}
\check{S}_{c,11} &=& \check{M} \check{Q}\check{M};\;
\check{M}= \left[\begin{array}{cc} \frac{\check{Z}}{\sqrt{2\cos(\theta)}} & 0 \cr 0 & \frac{\check{Z}^*}{\sqrt{2\cos(\theta^*)}}\end{array}\right];\; 
\check{Q}= \left[\begin{array}{cc} \check{Q}_{11} & \check{Q}_{12} \cr \check{Q}_{21} & \check{Q}_{22}\end{array}\right]\\
\check{Q}_{11} &=&\check{f} \check{s}_+ + (1-\check{f}) \check{s}_- + \check{Z}^{-2} \left(\check{f} \check{s}^T_+ + (1-\check{f}) \check{s}^T_-\right)\check{Z}^{-2} ;\\
\check{Q}_{12} &=&
 \left(- \check{f}\check{s}_+(1-\check{f}) + (1-\check{f}) \check{s}_- \check{f} \right) \check{Z}^{*-2} + \nonumber \\
 && \check{Z}^{-2} \left(- \check{f}\check{s}^T_+(1-\check{f}) + (1-\check{f}) \check{s}^T_- \check{f} \right) \\
\check{Q}_{21} &=&
 \left(\check{s}_- - \check{s}_+  \right) \check{Z}^{-2} + \check{Z}^{*-2} \left(\check{s}^T_- -\check{s}^T_+ \right) ;\; \\
\check{Q}_{22} &=&
 \check{s}^T_+(1-\check{f}) +\check{s}^T_-\check{f} +\check{Z}^{*-2} \left(\check{s}_+(1-\check{f}) +\check{s}_-\check{f}\right) \check{Z}^{*-2};\\
 \check{Q}_{12} = \check{Q}^T_{12};\; \check{Q}_{21} = \check{Q}^T_{21};\; \check{Q}_{11} = \check{Q}^T_{22}.
\end{eqnarray}
Since the determinant of $\check{M}$ does not depend on $\check{\chi}$, we can skip then and concentrate on ${\rm det} \check{Q}$,
\begin{equation}
{\cal S} = \frac{1}{2} \ln {\rm det} \check{Q}
\end{equation} 
This result provides the most complete description of the transport in a multi-terminal superconducting structure and is readily generalized to time- and energy dependent scattering matrix. Let us comprehend the structure of the answer  by contemplating several limits.

First of all, let us consider a limit of a non-superconducting circuit. Since ${\rm Im} \theta \to +\infty$ in this limit, we need to set $Z^{-1} \to 0$. This vanishes $\check{Q}_{12,21}$ and the determinant is a product of the determinants $\check{Q}_{11}$ and $\check{Q}_{22}$. The matrix $\check{Q}_{11}$ in this limit approaches 
$$
\check{Q}_{11} \to \check{f} \check{s}_+ + (1-\check{f}) \check{s}_- .
$$
This is equivalent to the matrix in Eq. \ref{eq:normalFCS}.  The matrix $\check{Q}_{22}$ in this limit approaches
$$
\check{Q}_{11} \to \check{s}^T_+(1-\check{f}) +\check{s}^T_-\check{f}
$$
which does not look the same. However, we note that $\check{Q}^T_{22} = \check{Q}_{11}$  if transposition includes $\epsilon \to -\epsilon$ (recall that we assume $f(\epsilon) = (1-f(-\epsilon)$). Since the determinant of a matrix is the same as that of the transposed one, ${\rm det}\check{Q}_{22}$ provides an equal contribution to the action. This double counting is removed by the $1/2$ prefactor in the action discussed.

Let us now derive the transport properties of the junction in the ground state. In this case, $f=\Theta(-\epsilon$, $1-f = \Theta(\epsilon)$, $f(1-f)$=0 so that $\check{Q}_{12}=0$ and again ${\rm det}\check{Q} = {\rm det}\check{Q}_{11} {\rm det}\check{Q}_{22} = ({\rm det}\check{Q}_{11})^2$ 
Let us assume for simplicity that $\theta$ is the same for all leads: this is the case of equal superconducting gaps in the leads. In this case,
\begin{eqnarray}
\check{Q}_{11} = \check{s}_- + e^{2 i \theta} \check{s}^T_-{\rm \ if \ \epsilon >0} \\
\check{Q}_{11} = \check{s}_+ + e^{2 i \theta} \check{s}^T_+{\rm \ if \epsilon >0}
\end{eqnarray} 
To proceed, we multiply the matrix with $\check{s}_-$ at positive $\epsilon$ and with $\check{s}_+$ at negative $\epsilon$. Since the determinant of these matrices does not depend on counting fields, this operation does not change the action of interest that now becomes
\begin{equation}
{\cal S} = {\cal T} \int \frac{d\epsilon}{2\pi} \left(
\Theta(\epsilon)\ln {\rm det}\left[1+ e^{2 i \theta}\check{s}^{-1}_-\check{s}^T_-\right] + \Theta(-\epsilon)
\ln {\rm det}\left[1+ e^{2 i \theta}\check{s}^{-1}_+\check{s}^T_+\right]
\right)
\end{equation} 
Now we can concentrate on the eigenvalues of the unitary matrices $s^{-1}_\pm\check{s}^T_\pm$, $\exp(i\lambda)$. The structure of the matrices is such that the eigenvalues form complex-conjugated pairs $\exp(i\pm \lambda)$. The eigenvalues $\exp(i\lambda_{\pm})$ of the matrices $s^{-1}_\pm\check{s}^T_\pm$ are obviously the functions of the superconducting phases $\{\phi_i\}$ shifted by counting fields. Let us now consider the integral 
$$
I(\lambda) = \int_0^{\infty} \frac{d\epsilon}{2\pi} \sum_{\pm} \left[1+ e^{2 i \theta}e^{i\pm\lambda}\right]
$$
Although this is not immediately obvious, the integral is purely imaginary. To prove this, we note 
that 
$$
I*(\lambda) = \int_0^{\infty} \frac{d\epsilon}{2\pi} \sum_{\pm} \left[1+ e^{-2 i \theta^*}e^{i\pm\lambda}\right] = \int_{-\infty}^0 \frac{d\epsilon}{2\pi} \sum_{\pm} \left[1+ e^{2 i \theta}e^{i\pm\lambda}\right]
$$
since $\theta(-\epsilon) = -\theta^*(\epsilon)$ and 
$$
I+I^* = \int_{-\infty}^{\infty} \frac{d\epsilon}{2\pi} \sum_{\pm} \left[1+ e^{2 i \theta}e^{i\pm\lambda}\right] = 0
$$
since the function  $e^{2 i \theta}$ is analytical in the upper half-plane of $\epsilon$ and the integration contour can be shifted there to infinity where the integral vanishes. 

With this, we are ready to represent the action as
\begin{equation}
{\cal S} = {\cal T} i (E_g(\{\phi_i +\chi_i\}) -E_g(\{\phi_i -\chi_i\}))
\end{equation}
$E_g$ being a phase-dependent part of the ground state energy of the junction. This is a form expected for FCS in a general ground state \cite{Kindermann} and in particular in superconductor\cite{Belzig}. The ground state energy is expessed
as a sum over the eigenvalues $\lambda(\{\phi_i\})$ of the $\lambda$-dependent part of the integral
$$
E_g = i \sum_{0<\lambda_i<\pi} I(\lambda_i)
$$
By substracting $\lambda$-independent part he integral can be transformed to 
\begin{equation}
\label{eq:Elambda}
-iI \equiv E_{\lambda}/2 = \int_0^{\infty} \frac{d\epsilon}{2\pi} {\rm arg} \left(\cos(2\theta)-\cos\lambda\right)
\end{equation} 
For purely BCS superconducting spectrum,
the argument is $\pi$ in the energy inverval $0<E<E_A$, where $E_A$ is the (positive) energy of a discrete Andreev 
bound state determined from the equation $2\theta(E_A)=\lambda$, and therefore $E_{\lambda}=E_A$. In more complex situations, the discrete bound state is not formed and/or the contribution to the energy can come from the states of the continous spectrum. In any case, every eigenvalue $\lambda$ contibutes to the energy with a term given by \ref{eq:Elambda}.

To conclude the section, we derive a useful formula for the superconducting currents averaged over statistical fluctuations. This relation is obtained by singling out from the action the terms proportional to the first power of $\chi$:
\begin{equation}
\check{I}  = -\frac{1}{2} {\rm diag} \left( [\check{s},\check{A}(1-2\check{f}) -(1-2\check{f}) \check{Z}^{*-2} \check{A}^T \check{Z}^{*-2}+ 2 \check{Z}^{-2} \check{A} \check{Q}^{(0)}_{12} \check{A}^T \right])
\end{equation}
where 
$$
 \check{A} = \left(\check{s} + \check{Z}^{-2} \check{s}^T\check{Z}^{-2} \right)^{-1};\; \check{Q}^{(0)}_{12} =[\check{s},\check{f}]\check{Z}^{*-2} +\check{Z}^2 [\check{f},\check{s}^T] 
$$
For a stationary case, the "check" structure in this formula is in the channel space, and it gives a contribution to the current in the channel $i$ at an energy. It needs to be integrated over energy to get the stationary current. For time-dependent situation, one includes time in the "check" structure so that the formula gives the current $I_i(t)$ while the energy-dependend $\check{\phi},\check{Z},\check{f}$ become the integral kernels in time.

\section{Conclusions}
\label{eq:conclusions}
In conclusion, we have provided a technical and comprehensive introduction to the Keldysh action formalism for a multi-terminal scatterer with special emphasis on superconducting leads. We have derived a very general and compact formula \ref{eq:answer} and have elaborated on simple important examples to demonstrate the variety of its applications.

I did this to commemorate Markus Buttiker, the pioneer of scattering approach to quantum transport, one of the fathers of this big, prosperous and fruitfully developing research field. I admire not only his research merits: throughout 25 years of our acquaintance I was appreciating much his daring to remain himself, to keep his own research style, research topics and idea sets in times where the close following of a quickly changing scientific fashion seemed to be a must. He was also a charming personality and a good friend.
  
\appendix
\section{Details of the variational integration}
Let us give here the details of the calculations between Eq. \ref{eq:variation} and Eq. \ref{eq:answer}. We work in the basis where 
\begin{equation}
\check{g} = \left[\begin{array}{cc} 1 & 0 \cr 0 &-1\end{array}\right]; \; \check{S} = \left[\begin{array}{cc} \check{S}_{11} & \check{S}_{12} \cr \check{S}_{21} & \check{S}_{22}\end{array}\right]
\end{equation}
The variation of $\check{g}$ anticommutes with $\check{g}$ and in this basis generally reads
\begin{equation}
\check{g} = \left[\begin{array}{cc} 0 & \check{V} \cr \check{W} & 0\end{array}\right]
\end{equation}
The inverse matrix playing important role becomes
\begin{equation}
\frac{1}{\check{g} \check{S} + \check{S}\check{g}} = 
\left[\begin{array}{cc} \check{S}^{-1}_{11} & 0 \cr 0 & -\check{S}^{-1}_{22}\end{array}\right]
\end{equation}
With this, the variation of action becomes
\begin{equation}
\delta {\cal S} = {\rm Tr} \left[ \check{S}_{21} \check{W} \check{S}^{-1}_{11} +  \check{V} \check{S}_{12} \check{S}^{-1}_{11}\right]
\end{equation}
Let us now determine how the blocks of $\check{S}$ are transformed 
upon the variation of $\check{g}$. The matrix transforming $\check{G}$ back to the diagonal form reads 
\begin{equation}
\check{L},\check{L}^{-1} = 1 \pm \frac{1}{2} \left[\begin{array}{cc} 0 & \check{V} \cr -\check{W} & 0\end{array}\right]
\end{equation}
Applying this to $\check{S}$, $\check{S} \to \check{L} \check{S} \check{L}^{-1}$, we observe that
$$
\delta \check{S}_{11} = \check{V} \check{S}_{21} + \check{S}_{12} \check{W}.
$$
Therefore,
\begin{equation}
\delta {\cal S} = {\rm Tr} \left[\delta \check{S}_{11} \check{S}_{11}^{-1} \right]
\end{equation}
which makes the integration straightforward.


\begin{thebibliography}{99}
\bibitem{AGD} A. A. Abrikosov, L. P. Gorʹkov, and I. E. Dzyaloshinski, {\it Methods of Quantum Field Theory in Statistical Physics}, 
Courier Corporation, 1975. 
\bibitem{Landavshiz} L. P. Pitaevskii and E.M. Lifshitz, {\it Physical Kinetics},
Butterworth-Heinemann, 2012.
\bibitem{LB1} M. Buttiker, Y. Imry, R. Landauer, and S. Pinhas, Phys. Rev. B {\bf 31}, 6207 (1985).
\bibitem{LB2} M Buttiker, Phys. Rev. Lett. {\bf 57}, 1761 (1986).
\bibitem{NoiseButtiker} M. Buttiker
Phys. Rev. Lett. {\bf 65}, 2901 (1990); Y.M. Blanter and M. Buttiker
Phys. Rep. {\bf 336}, 1 (2000).
\bibitem{NoiseLesovik} G.B. Lesovik, JETP Lett., {\bf 49} , 592 (1989).
\bibitem{LesovikLevitov} L.S. Levitov, G.B. Lesovik,  JETP Lett. {\bf 58} , 230 (1993).
\bibitem{QT} Yuli V. Nazarov
\bibitem{BB} C.W.J Beenakker and M. Buttiker
Physical Review B {\bf 46}, 1889 (1992).
\bibitem{Ohm} Yuli V. Nazarov,  in:{\it Quantum Dynamics of Submicron Structures} · NATO ASI Series Volume {\bf 291}, 687 (1995).
\bibitem{Efetov} I. S. Beloborodov, K. B. Efetov, A. V. Lopatin, V. M. Vinokur, Rev. Mod. Phys. {\bf 79}, 469 (2007).
\bibitem{NazarovBagrets} Yu. V. Nazarov and D. A. Bagrets, Phys. Rev. Lett. {\bf 88}, 196801 (2002). 
\bibitem{MeGrenoble} R.-P. Riwar, M. Houzet, J. S. Meyer, and Yu. V. Nazarov, arXiv:1503.06862.
\bibitem{Izak} I. Snyman, Y. V. Nazarov
Phys. Rev. B {\bf 77}, 165118 (2008).
\bibitem{Kindermann} Yu. V. Nazarov, M. Kindermann
 Eur. Phys. J. B {\bf 35}, 413 (2003). 
 \bibitem{Belzig} W. Belzig and Yu. V. Nazarov,   Phys. Rev. Lett. {\bf 87}, 197006 (2001). 
\end{thebibliography}
\end{document}